\documentclass[apjl]{emulateapj}
\usepackage{comment}

\newcommand{\ctbd}[1]{}

\newcommand{\lc}{light curve}
\newcommand{\lcs}{light curves}
\newcommand{\Lc}{Light curve}

\newcommand{\kms}{\ensuremath{\rm km\,s^{-1}}}
\newcommand{\ms}{\ensuremath{\rm m\,s^{-1}}}

\newcommand{\gcmc}{\ensuremath{\rm g\,cm^{-3}}}
\newcommand{\ergs}{\ensuremath{\rm erg\,s^{-1}\,cm^{-2}}}

\newcommand{\hd}[1]{\mbox{HD #1}}

\newcommand{\vsini}{\ensuremath{v \sin{i}}}

\newcommand{\rsun}{\ensuremath{R_\sun}}
\newcommand{\msun}{\ensuremath{M_\sun}}

\newcommand{\rstar}{\ensuremath{R_\star}}

\newcommand{\teffstar}{\ensuremath{T_{\rm eff}}}

\newcommand{\mearth}{\ensuremath{M_\earth}}

\newcommand{\mpl}{\ensuremath{M_{p}}}

\newcommand{\rjup}{\ensuremath{R_{\rm J}}}
\newcommand{\mjup}{\ensuremath{M_{\rm J}}}

\newcommand{\figr}[1]{Fig.~\ref{fig:#1}}
\newcommand{\secr}[1]{\mbox{\S\ \ref{sec:#1}}}

\newcommand{\tabr}[1]{\mbox{Table~\ref{tab:#1}}}

\newcommand{\flwof}{\mbox{FLWO 1.2 m}}

\newcommand{\flwos}{\mbox{FLWO 1.5 m}}

\newcommand{\band}[1]{\ensuremath{#1}-band}

\newcommand{\hatcur}{HAT-P-10}
\newcommand{\hatcurb}{HAT-P-10b}

\newcommand{\hatcurCCra}{03h09m28.55s}                                 
\newcommand{\hatcurCCdec}{+30d40m24.9s}                                
\newcommand{\hatcurCCtwomass}{2MASS~03092855+3040249}                  
\newcommand{\hatcurCCgsc}{GSC~02340-01714}                             
\newcommand{\hatcurCCtassmv}{11.89}                                    
\newcommand{\hatcurCCtassvi}{\ensuremath{1.09\pm0.11}}                 
\newcommand{\hatcurLCdip}{\ensuremath{15}}                             
\newcommand{\hatcurLCrprstar}{\ensuremath{0.1332\pm0.0013}}            
\newcommand{\hatcurLCimp}{\ensuremath{0.238_{-0.093}^{+0.130}}}        
\newcommand{\hatcurLCdur}{\ensuremath{0.1100\pm0.0015}}                
\newcommand{\hatcurLCingdur}{\ensuremath{0.0136\pm0.0014}}             
\newcommand{\hatcurLCP}{\ensuremath{3.7224690\pm0.0000067}}            
\newcommand{\hatcurLCPprec}{\ensuremath{3.7224690}}                    
\newcommand{\hatcurLCPshort}{\ensuremath{3.7225}}                      
\newcommand{\hatcurLCT}{\ensuremath{2454729.90631\pm0.00030}}          
\newcommand{\hatcurSMEteff}{\ensuremath{4980\pm60}}                    
\newcommand{\hatcurSMEzfeh}{\ensuremath{0.13\pm0.08}}                  
\newcommand{\hatcurSMElogg}{\ensuremath{4.5\pm0.1}}                    
\newcommand{\hatcurSMEvsin}{\ensuremath{0.5\pm0.2}}                    
\newcommand{\hatcurYYm}{\ensuremath{0.82\pm0.03}}                      
\newcommand{\hatcurYYmshort}{\ensuremath{0.82}}                        
\newcommand{\hatcurYYr}{\ensuremath{0.81_{-0.02}^{+0.03}}}             
\newcommand{\hatcurYYrshort}{\ensuremath{0.81}}                        
\newcommand{\hatcurYYlogg}{\ensuremath{4.54\pm0.03}}                   
\newcommand{\hatcurYYlum}{\ensuremath{0.36_{-0.03}^{+0.04}}}           
\newcommand{\hatcurYYmv}{\ensuremath{6.12\pm0.12}}                     
\newcommand{\hatcurYYvi}{\ensuremath{0.917\pm0.019}}                   
\newcommand{\hatcurYYage}{\ensuremath{11.2\pm4.1}}                     
\newcommand{\hatcurRVK}{\ensuremath{69.1\pm3.5}}                       
\newcommand{\hatcurRVgamma}{\ensuremath{35.5\pm3.0}}                   
\newcommand{\hatcurPPi}{\ensuremath{88.5\pm0.6}}                       
\newcommand{\hatcurPPlogg}{\ensuremath{3.02_{-0.05}^{+0.04}}}          
\newcommand{\hatcurPPar}{\ensuremath{11.93_{-0.57}^{+0.21}}}           
\newcommand{\hatcurPParel}{\ensuremath{0.0439_{-0.0009}^{+0.0006}}}    
\newcommand{\hatcurPPrho}{\ensuremath{0.498\pm0.064}}                  
\newcommand{\hatcurPPm}{\ensuremath{0.46\pm0.03}}                      
\newcommand{\hatcurPPmlong}{\ensuremath{0.460\pm0.028}}                
\newcommand{\hatcurPPr}{\ensuremath{1.05_{-0.03}^{+0.05}}}             
\newcommand{\hatcurPPrshort}{\ensuremath{1.05}}                        
\newcommand{\hatcurPPrlong}{\ensuremath{1.045_{-0.033}^{+0.050}}}      
\newcommand{\hatcurPPmrcorr}{\ensuremath{0.025}}                       
\newcommand{\hatcurPPteff}{\ensuremath{1030_{-19}^{+26}}}              
\newcommand{\hatcurPPtheta}{\ensuremath{0.047\pm0.003}}                
\newcommand{\hatcurXdist}{\ensuremath{125_{-5}^{+6}}}                  

\newcommand{\hatcurYYlumshort}{\ensuremath{0.36}}                      
\renewcommand{\hatcurCCra}{\ensuremath{03^{\mathrm h}09^{\mathrm m}28.55^{\mathrm s}}}  
\renewcommand{\hatcurCCdec}{ \ensuremath{+30^{\mathrm d}40^{\mathrm m}24.9^{\mathrm s}} }                                

\shortauthors{Bakos et al.}
\shorttitle{HAT-P-10b}

\begin{document}

\title{\hatcur\lowercase{b}: A light and moderately hot Jupiter transiting
	a K dwarf\altaffilmark{$\dagger$}}

\author{
	G.~\'A.~Bakos\altaffilmark{1,2},
	A.~P\'al\altaffilmark{1,4},
	G.~Torres\altaffilmark{1},
	B.~Sip\H{o}cz\altaffilmark{1,4},
	D.~W.~Latham\altaffilmark{1},
	R.~W.~Noyes\altaffilmark{1},
	G\'eza~Kov\'acs\altaffilmark{3},
	J.~Hartman\altaffilmark{1},
	G.~A.~Esquerdo\altaffilmark{1},
	D.~A.~Fischer\altaffilmark{6},
	J.~A.~Johnson\altaffilmark{7},
	G.~W.~Marcy\altaffilmark{5},
	R.~P.~Butler\altaffilmark{8},
	A.~Howard\altaffilmark{5},
	D.~D.~Sasselov\altaffilmark{1},
	G\'abor~Kov\'acs\altaffilmark{1},
	R.~P.~Stefanik\altaffilmark{1},
	J.~L\'az\'ar\altaffilmark{9},
	I.~Papp\altaffilmark{9},
	P.~S\'ari\altaffilmark{9}
}
\altaffiltext{1}{Harvard-Smithsonian Center for Astrophysics,
	Cambridge, MA, gbakos@cfa.harvard.edu}

\altaffiltext{2}{NSF Fellow}

\altaffiltext{3}{Konkoly Observatory, Budapest, Hungary}

\altaffiltext{4}{Department of Astronomy,
	E\"otv\"os Lor\'and University, Budapest, Hungary.}

\altaffiltext{5}{Department of Astronomy, University of California,
	Berkeley, CA}

\altaffiltext{6}{Department of Physics and Astronomy, San Francisco
	State University, San Francisco, CA}

\altaffiltext{7}{Institute for Astronomy, University of Hawaii,
Honolulu, HI 96822; NSF Postdoctoral Fellow}

\altaffiltext{8}{Department of Terrestrial Magnetism, Carnegie
	Institute of Washington, DC}

\altaffiltext{9}{Hungarian Astronomical Association, Budapest, 
	Hungary}

\altaffiltext{$\dagger$}{%
	Based in part on observations obtained at the W.~M.~Keck
	Observatory, which is operated by the University of California and
	the California Institute of Technology. Keck time has been
	granted by NOAO (A285Hr) and NASA (N128Hr).
}


\begin{abstract} 

We report on the discovery of \hatcurb{}, the lowest mass
(\hatcurPPm\,\mjup) transiting extrasolar planet (TEP) discovered to
date by transit searches. \hatcurb\ orbits the moderately bright
V=\hatcurCCtassmv\ K dwarf \hatcurCCgsc, with a period
$P=\hatcurLCP\,d$, transit epoch $T_c = \hatcurLCT$ (BJD) and duration
\hatcurLCdur\,d. \hatcurb\ has a radius of \hatcurPPr\,\rjup\ yielding
a mean density of \hatcurPPrho\,\gcmc. Comparing these observations
with recent theoretical models we find that \hatcur\ is consistent with
a $\sim 4.5\,{\rm Gyr}$, coreless, pure hydrogen and helium gas giant
planet. With an equilibrium temperature of $T_{eq} = \hatcurPPteff\,K$,
\hatcurb\ is one of the coldest TEPs. Curiously, its Safronov number
$\theta=\hatcurPPtheta$ falls close to the dividing line between the
two suggested TEP populations.

\end{abstract}



\section{Introduction}
\label{sec:introduction}

It has become clear in recent years that transiting extrasolar planets
(TEPs), especially those around bright stars, are extremely valuable
for understanding the physical properties of planetary bodies. The
transit itself is a periodic event, which --- together with high
precision spectroscopic observations and radial velocity (RV) follow-up
-- reveals a number of key parameters, notably the relative radius of
the planet with respect to the star, and the true mass of the planet
without the inclination ambiguity. These allow determination of the
mean density of the planet, and an insight into its basic structural
properties. These advantages have been realized early on, and the
recent rise in the detection of TEPs is due to a number of dedicated
transit searches, such as TrES \citep{brown:00,dunham:04}, XO
\citep{pmcc:05}, SuperWASP \citep{pollacco:06}, OGLE \citep[targeting
fainter stars;][]{udalski:08}, and HATNet \citep{bakos:02,bakos:04}. 
At the time of this writing, the number of published TEPs with a unique
identification is $\sim 40$, with $\sim 35$ of these due to systematic
searches. The properties of known TEPs already span a wide range, from
the hot Neptune GJ436b with a mass of $\mpl = 0.072\,\mjup$
\citep{butler:04,gillon:07} to XO-3 with $\mpl = 11.79\mpl$
\citep{johns-krull:08}, from short period orbits like OGLE-TR-56b with
$P=1.2$\,days \citep{udalski:02,konacki:03} to $P=21.2$\,days of
\hd{17506}b \citep{barbieri:07}. Although most of these planets have
circular orbits, some planets with significant eccentricities, such as
HAT-P-2b \citep{bakos:07a}, have also been reported. TEPs have been
discovered in a wide range of environments, from orbiting M dwarfs
(GJ436b) up to mid F-dwarfs, such as HAT-P-7b \citep{pal:08a}.

Theoretical investigations have been thriving during this vigorous
discovery era, some focusing on the radius of these planets
\citep[][]{burrows:07,liu:08,chabrier:04,fortney:07}, and others on the
atmospheres \citep[e.g.][]{burrows:06,fortney:07}, to mention two of
the key observable properties of transiting planets. When confronting
theory with observations, it is also essential to use accurate
observational values, along with proper error estimates. The recent
compilation of TEP parameters by \citet{torres:08} represents a step
forward in this sense. It was also noted throughout these works that a
much larger sample is required for better understanding of the
underlying physics, i.e.~more planets are needed to populate the
mass--radius (or other) parameter space, to improve the statistical
significance of correlations between planetary and stellar parameters,
and to reveal any previously undetected correlations that may shed
light on the physical processes governing the formation and evolution
of TEPs.

The HATNet survey has been a major contributor to TEP discoveries.
Operational since 2003, it has covered approximately 7\% of the
Northern sky, searching for TEPs around bright stars ($8\lesssim I
\lesssim 12$\,mag). HATNet operates six wide field instruments: four at
the Fred Lawrence Whipple Observatory (FLWO) in Arizona, and two on the
roof of the Submillimeter Array Hangar (SMA) of SAO\@. Since 2006,
HATNet has announced and published 9 TEPs. In this work we report on
the tenth such discovery.\footnote{After submission of this paper, it
was realized that HAT-P-10b and WASP-11b refer to the same object, 
independently discovered, with WASP-11b submitted 7 days earlier to
A\&A. The two discovery groups agreed on calling it in future papers 
as WASP-11b/HAT-P-10b, with separate entries on www.exoplanet.eu}.

\section{Photometric detection}
\label{sec:detection}

\begin{figure}[!ht]
\plotone{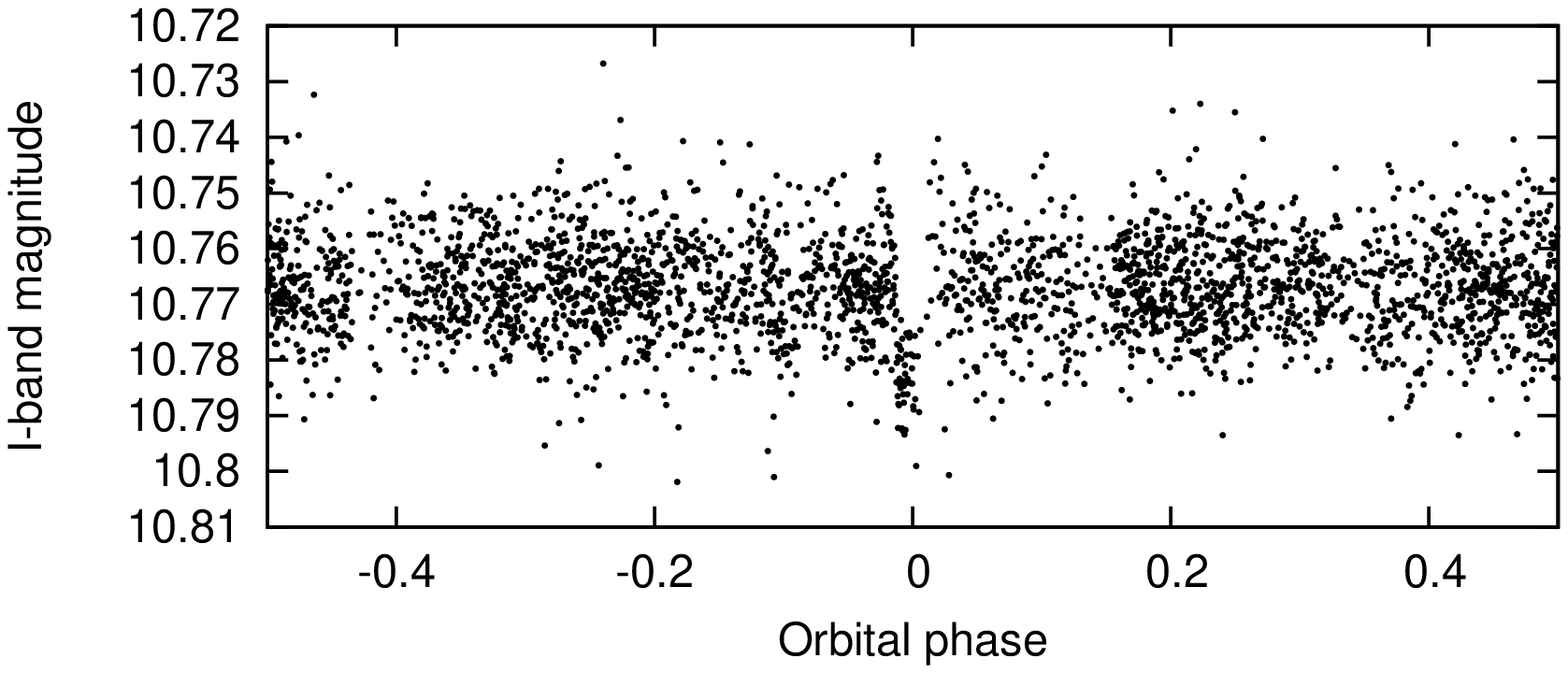}
\caption{
	The unbinned \lc{} of \hatcur{} including all 2870
	instrumental \band{I} measurements obtained with 
	the HAT-10 telescope of HATNet (see text for details), and folded 
	with the period of $P = \hatcurLCPprec$\,days 
	(which is the result the fit described in \secr{analysis}). 
\label{fig:hatnet}}
\end{figure}

The transits of \hatcurb{} were detected with one of the HATNet
telescopes, \mbox{HAT-10}, located at FLWO\@. The region around
\hatcurCCgsc{}, a field internally labeled as G213, was observed on a
nightly basis between 2005 October 3 and 2006 March 14, whenever
weather conditions permitted. We gathered 2870 exposures of 5 minutes
at a 5.5-minute cadence. Each image contained approximately $29,000$
stars down to $I\sim14.0$. For the brightest stars in the field we
achieved a per-image photometric precision of 4\,mmag.

The calibration of the HATNet frames was done utilizing standard
procedures. The calibrated frames were then subject to star detection
and astrometry, as described in \cite{pal:06}. Aperture photometry was
performed on each image at the stellar centroids derived from the 2MASS
catalog \citep{cutri:03} and the individual astrometrical solutions.
The resulting \lcs\ were decorrelated against trends using the External
Parameter Decorrelation technique \citep[EPD, see][]{bakos:07b} and the
Trend Filtering Algorithm \citep[TFA, see][]{kovacs:05}. The \lcs{}
were searched for periodic box-like signals using the Box Least Squares
method \citep[BLS, see][]{kovacs:02}. We detected a significant signal
in the \lc{} of \hatcurCCgsc{} (also known as \hatcurCCtwomass{};
$\alpha = \hatcurCCra$, $\delta = \hatcurCCdec$; J2000), with a depth
of $\sim\hatcurLCdip$\,mmag, and a period of $P=\hatcurLCPshort$\,days.
The dip had a relative duration (first to last contact) of
$q\approx0.027$, equivalent to a total duration of $Pq\approx2.5$~hours
(see \figr{hatnet}).


\section{Follow-up observations}
\label{sec:followup}

\subsection{Reconnaissance Spectroscopy}

All HATNet candidates are subject to thorough investigation before
using more precious time on large telescopes, such as Keck I, to
observe them. One of the important tools for establishing whether the
transit-feature in the light curve of a candidate is due to
astrophysical phenomena other than a planet transiting a star is the
CfA Digital Speedometer \citep[DS;][]{latham:92}, mounted on the
\flwos\ telescope.

High-resolution spectra with low signal-to-noise ratio from this
facility have been used routinely in the past to derive radial
velocities with moderate precision (roughly 1\,\kms) and to classify
the effective temperature and surface gravity of the host star, to weed
out false alarms, such as F dwarfs orbited by M dwarfs, grazing
eclipsing binaries, triple and quadruple star systems, or giant stars
where the transit signal is either false, or comes from a nearby,
blended eclipsing binary.

The RV measurements of \hatcur{} showed an rms residual of 0.43\,\kms,
consistent with no detectable RV variation. Atmospheric parameters for
the star, including the effective temperature $T_{\rm eff} = 5000\,K$,
surface gravity $\log g = 4.5$, and projected rotational velocity
$\vsini = 1.5\kms$, were derived as described by \cite{torres:02}. The
effective temperature and surface gravity corresponds to an early K
dwarf.

\subsection{High resolution, high S/N spectroscopy}

Given the significant detection by HATNet, and the positive DS results
that exclude the usual suspects, we proceeded with the follow-up of
this candidate by obtaining high-resolution and high S/N spectra to
characterize the radial velocity variations and to determine the
stellar parameters with higher precision. We obtained 6 exposures with
an iodine cell, plus one iodine-free template, using the HIRES
instrument \citep{vogt:94} on the Keck~I telescope located on Mauna
Kea, Hawaii. The observations were made on the nights of 2008 March
21-22, July 27 and on three nights between September 13 and 17. The
small RV variations based on the March 2007 run made this target a firm
planet candidate, but more observations were required to derive an
orbit, and to check spectral bisector variations (see \secr{blend}).

\begin{figure} [ht]
\plotone{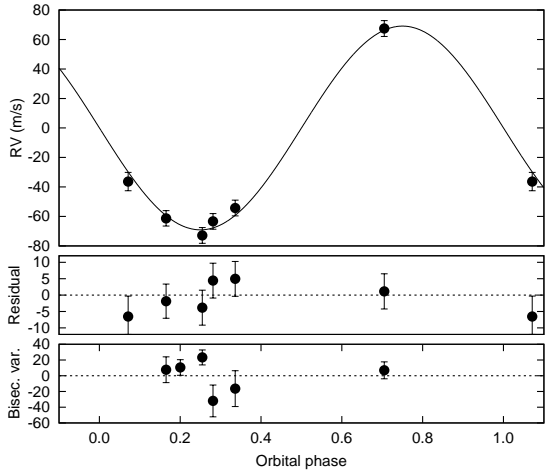}
\caption{
	(Top) Radial-velocity measurements from Keck for \hatcur{}, along
	with an orbital fit, shown as a function of orbital phase, using
	our best fit period (see \secr{analysis}). The center-of-mass
	velocity has been subtracted.
	(Middle) Phased residuals after subtracting the orbital fit (also
	see \secr{analysis}). The rms variation of the residuals is about
	$4.2$\,\ms.
	(Bottom) Bisector spans (BS) for 5 of the 6 Keck spectra plus the
	single template spectrum (\secr{blend}).  The mean value has been
	subtracted.  Note the different vertical scale of the panels.
\label{fig:rvbis}}
\end{figure}

The width of the spectrometer slit used on HIRES was $0\farcs86$,
resulting in a resolving power of $\lambda/\Delta\lambda \approx
55,\!000$, with a wavelength coverage of $\sim3800-8000$\,\AA\@. The
iodine gas absorption cell was used to superimpose a dense forest of
$\mathrm{I}_2$ lines on the stellar spectrum and establish an accurate
wavelength fiducial \citep[see][]{marcy:92}. Relative RVs in the Solar
System barycentric frame were derived as described by \cite{butler:96},
incorporating full modeling of the spatial and temporal variations of
the instrumental profile. The final RV data and their errors are listed
in \tabr{rvs}. The folded data, with our best fit (see \secr{analysis})
superimposed, are plotted in \figr{rvbis}.

\begin{deluxetable}{lrrc}
\tablewidth{0pc}
\tablecaption{Relative radial velocity measurements 
of \hatcur{}\label{tab:rvs}}
\tablehead{
	\colhead{BJD} & 
	\colhead{RV} & 
	\colhead{\ensuremath{\sigma_{\rm RV}}
}\\
\colhead{
	\hbox{~~~~(2,454,000$+$)~~~~}} & 
	\colhead{(\ms)} & 
	\colhead{(\ms)}
}
\startdata
547.76983    \dotfill &  $     -0.9$ &   $       3.7$  \\
548.75614    \dotfill &  $    -18.8$ &   $       1.8$  \\
675.11598    \dotfill &  $    -27.8$ &   $       1.8$  \\
723.07574    \dotfill &  $    -25.8$ &   $       1.5$  \\
725.08480    \dotfill &  $    103.0$ &   $       2.0$  \\
727.13146    \dotfill &  $    -37.4$ &   $       1.8$ 
\enddata
\end{deluxetable}

\subsection{Photometric follow-up observations}
\label{sec:phot}

\begin{figure}[!ht]
\plotone{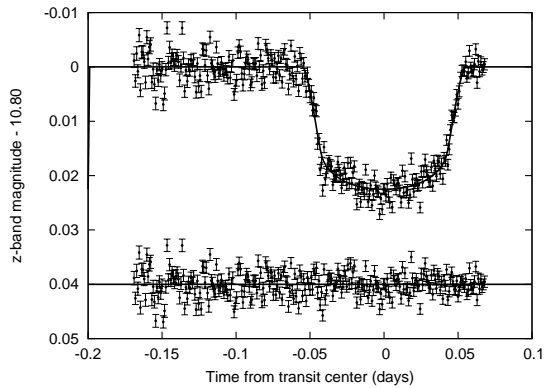}
\caption{
	Unbinned instrumental Sloan \band{z} transit \lc\ acquired with
	KeplerCam at the \flwof{} telescope on 2008 September 19 MST;
	superimposed is the best-fit transit model \lc{}.  Below we show
	the residuals from the fit.
\label{fig:lc}}
\end{figure}

We observed a complete transit event of \hatcur{} on the night of 2008
September 19/20 MST with the KeplerCam CCD on the \flwof\ telescope.
Altogether $278$ frames were acquired with a cadence of $65$ seconds in
Sloan $z$ band. The reduction of the images was performed as follows.
After bias and flat calibration, we derived an initial second order
astrometrical transformation between the $\sim110$ brightest stars and
the 2MASS catalog, as described in \citet{pal:06}, yielding a residual
of $\sim0.3$ pixels. In order to avoid systematic errors resulting from
the proper motion of the stars, we generated a new catalog. This
catalog was based on the detected stellar centroids, the coordinates of
which were transformed to the same reference system using the initial
astrometrical solutions, and then averaged out using 3-$\sigma$
rejection. Using this new catalogue as reference, the final
astrometrical solution was derived for each frame, yielding a residual
of $\sim0.04$ pixel. In the next step, aperture photometry was
performed using a series of apertures with radii of $6.0$, $7.5$, $9.0$
and $10.5$ pixels. The instrumental magnitude transformation was also
done in two steps: first, all magnitude values were transformed to the
photometric reference frame (selected to be the sharpest image), using
the individual Poisson noise error estimations as weights. In the
second step, the magnitude fit was repeated using the mean individual
\lc{} magnitudes as reference and the rms of these \lcs{} as weights.
In both of the magnitude transformations, we excluded from the fit the
target star itself and the $3$-$\sigma$ outliers. We performed EPD
against trends simultaneously with the light curve modeling (for more
details, see \secr{analysis}). From the series of apertures we chose
the one with a radius of $9.0$ pixels, yielding the smallest fit rms.
This aperture falls in the middle of the aperture series, confirming
the plausible selection for the apertures.  The final \lc{} is shown in
the upper plot of \figr{lc}, superimposed with our best fit transit
\lc{} model (see also \secr{analysis}).


\section{Analysis}
\label{sec:analysis}

In this section we describe briefly our analysis yielding the orbital,
planetary and stellar parameters of the \hatcur\ system.

\subsection{Planetary, orbital and stellar parameters}

First, using the template spectrum obtained by the Keck/HIRES
instrument, we derived the stellar atmospheric parameters. We used the
SME package of \cite{valenti:96}, which yielded the following values
with conservative errors:
$\teffstar=\hatcurSMEteff$\,K, 
$\log g_\star=\hatcurSMElogg$\,(cgs),
$\mathrm{[Fe/H]}=\hatcurSMEzfeh$, and 
$v\sin i=\hatcurSMEvsin\,\mathrm{km\,s^{-1}}$. 

In modeling both the HATNet and the follow-up transit light curves, we
used the quadratic limb darkening formalism of \cite{mandel:02}. The
limb darkening coefficients used for the above stellar atmospheric
parameters by interpolating in the tables provided by \cite{claret:04}.
The coefficients we derived for $I$ and $z$ photometric passbands were
$\gamma_{1(I)}=0.3806$, $\gamma_{2(I)}=0.2535$, 
$\gamma_{1(z)}=0.3214$, and $\gamma_{2(z)}=0.2693$.

Following this, a joint fit was done using all of the available data,
including the HATNet \lc{}, the follow-up \lc{} and the radial velocity
measurements. Throughout the analysis, we refer to the transit event
observed on 2008 September 19/20 as $N_{\rm tr}=0$.

We adjusted the following parameters:
$T_{\mathrm{c},-290}$, the time of first transit center in the HATNet
campaign;
$T_{\mathrm{c},0}$, the time of the transit center on September 19/20; 
$K$, the radial velocity semi-amplitude; 
$k=e\cos\omega$ and $h=e\sin\omega$, the Lagrangian orbital elements
related to the eccentricity and argument of periastron;
$p\equiv R_{\rm p}/\rstar$, the fractional planetary radius; 
$b^2$, the square of the impact parameter;
the quantity $\zeta/\rstar$, which is related 
to the transit duration $T_{\rm dur}$ as 
$(\zeta/\rstar)^{-1}=T_{\rm dur}/2$;
and $M_{0}$ and $M_{1}$, the out-of-transit instrumental magnitudes of
the HATNet and FLWO/KeplerCam \lcs{}.
As noted by \cite{bakos:07b}, the quantity $\zeta/\rstar$ shows only a
small correlation with the other light curve parameters ($R_{\rm
p}/\rstar$, $b^2$), which makes it a good parameter to use. For
eccentric orbits, this quantity is related to the normalized semi-major
axis $a/\rstar$ as
$\zeta/\rstar=(2\pi/P)(a/\rstar)\sqrt{1-e^2}(1-b^2)^{-1/2}(1+h)^{-1}$.
To find the best fit values and the uncertainties, we utilized the
method of Markov Chain Monte-Carlo \citep[MCMC;][]{ford:06} which
provides the \emph{a posteriori} distribution of the adjusted
parameters.

The values and uncertainties of the $k$ and $h$ orbital elements were
found to be consistent with zero within $1$-$\sigma$, namely
$k=0.04\pm0.11$ and $h=0.11\pm0.12$. Therefore we conclude that the
observations are consistent with a circular planetary orbit, and we
repeated the fit by fixing the eccentricity to zero.

The results for the simultaneous fit are reported in \tabr{parameters},
except for the auxiliary parameters
$T_{\mathrm{c},-290}=2453650.39029\pm0.00195$~(BJD), 
$T_{\mathrm{c},0}=\hatcurLCT$~(BJD) that are used to derive the epoch
and the period as shown in \cite{pal:08a}. The RV jitter is the
additional velocity uncertainty that should be added quadratically to
the nominal errors (estimated from the Poisson-noise) in order to have
a reduced $\chi^2$ of unity (this is the quadratic sum of the
residuals, divided by the degrees of freedom of the RV fit, i.e.~4).
Our final value for the jitter is $4.2\,\ms$, and the error-bars on
\figr{rvbis} (top and middle panel) have been inflated accordingly.

\begin{deluxetable}{lcl}
\tablewidth{0pc}
\tablecaption{Stellar parameters for \hatcur{} \label{tab:stellar}}
\tablehead{\colhead{Parameter}	& \colhead{Value} 		& \colhead{Source}}
\startdata
$\teffstar$ (K)\dotfill			&  \hatcurSMEteff	& SME\tablenotemark{a} 		\\
$[\mathrm{Fe/H}]$\dotfill		&  \hatcurSMEzfeh	& SME						\\
$v \sin i$ (\kms)\dotfill		&  \hatcurSMEvsin	& SME						\\
$M_\star$ ($M_{\sun}$)\dotfill  &  \hatcurYYm		& Y$^2$+LC+SME\tablenotemark{b}	\\
$\rstar$ ($R_{\sun}$)\dotfill	&  \hatcurYYr		& Y$^2$+LC+SME			\\
$\log g_\star$ (cgs)\dotfill    &  \hatcurYYlogg	& Y$^2$+LC+SME			\\
$L_\star$ ($L_{\sun}$)\dotfill  &  \hatcurYYlum		& Y$^2$+LC+SME			\\
$M_V$ (mag)\dotfill				&  \hatcurYYmv   	& Y$^2$+LC+SME			\\
Age (Gyr)\dotfill				&  \hatcurYYage		& Y$^2$+LC+SME			\\
Distance (pc)\dotfill			&  \hatcurXdist		& Y$^2$+LC+SME
\enddata
\tablenotetext{a}{SME = `Spectroscopy Made Easy' package for analysis
of high-resolution spectra \cite{valenti:96}. See text.}
\tablenotetext{b}{Y$^2$+LC+SME = Yale-Yonsei 
isochrones \citep{yi:01}, \lc{} parameters, and SME results.}
\end{deluxetable}

The results of the joint fit, together with the initial results from
spectroscopy enable us to refine the parameters of the star.  As
described by \cite{sozzetti:07} and \cite{torres:08}, $a/\rstar$ is a
better luminosity indicator than the spectroscopic value of $\log
g_\star$ since stellar surface gravity has only a subtle effect on the
line profiles. Therefore, we used the values of $\teffstar$ and
$\mathrm{[Fe/H]}$ from the initial SME analysis, together with the
distribution of $a/\rstar$ to estimate the stellar properties from
comparison with the Yonsei-Yale (Y$^2$) stellar evolution models by
\cite{yi:01} and \cite{demarque:04}. Using the relation between
$a/\rstar$ and $\zeta/\rstar$, we derive the \emph{a posteriori}
distribution for the former one, and used the derived stellar density
as an input for the stellar evolution models in order to have an
\emph{a posteriori} distribution for the stellar parameters
\citep[see][for more details]{pal:08a,pal:08c}. Since the mass and
radius (and their respective distributions) of the star are known, it
is straightforward to obtain the surface gravity and its uncertainty
together. The derived surface gravity is $\log g_\star=\hatcurYYlogg$.
Since the surface gravity from the initial SME analysis agrees well
with the one derived above, we accept the latter as final value
(listed, together with other parameters in \tabr{stellar}).

The stellar evolution modeling also yields the absolute magnitudes and
colors in various photometric passbands. The derived $V-I$ color of the
star is $(V-I)_{\rm YY}=\hatcurYYvi$, slightly smaller than the color
from the TASS catalog \citep{droege:06}, namely $(V-I)_{\rm
TASS}=\hatcurCCtassvi$. Since this excess is most likely due to
interstellar reddening, we used the 2MASS $J$ magnitude to estimate the
distance. The observed $J$ band magnitude of \hatcur{} is
$J=10.015\pm0.020$ while the stellar modeling gives
$M_{J}=4.530\pm0.087$, which lead to a distance modulus for the star of
$J-M_{J}=5.484\pm0.089$, corresponding to a distance of
$d=\hatcurXdist$\,pc.

The planetary parameters and their uncertainties can be derived by
direct combination of the \emph{a posteriori} distributions of the
light curve, radial velocity and stellar parameters \citep[see
also][]{pal:08a}. We find that the mass of the planet is
$M_p=\hatcurPPmlong$\,\mjup, the radius is $R_p=\hatcurPPrlong$\,\rjup,
and its density is $\rho_p=\hatcurPPrho$\,\gcmc. The final planetary
parameters are summarized at the bottom of \tabr{parameters}.

\begin{deluxetable}{lc}
\tablewidth{0pc}
\tablecaption{Orbital and planetary parameters\label{tab:parameters}}
\tablehead{\colhead{~~~~~~~~~~~~~~~Parameter~~~~~~~~~~~~~~~} & \colhead{Value}}
\startdata
\sidehead{\Lc{} parameters}
~~~$P$ (days)			\dotfill 		& $\hatcurLCP$ 		\\
~~~$E$ (${\rm BJD}$)		\dotfill	& $\hatcurLCT$		\\
~~~$T_{14}$ (days)
	\tablenotemark{a} 	\dotfill		& $\hatcurLCdur$	\\
~~~$T_{12} = T_{34}$ (days)
	\tablenotemark{a} 	\dotfill		& $\hatcurLCingdur$	\\
~~~$b^2$                \dotfill    	& $0.092\pm0.062$ \\
~~~$\zeta/\rstar$\,$\mathrm{day^{-1}}$ \dotfill    & $20.72\pm0.14$\\
~~~$a/\rstar$			\dotfill        & $\hatcurPPar$		\\
~~~$R_p/\rstar$		\dotfill        	& $\hatcurLCrprstar$	\\
~~~$b \equiv a \cos i/\rstar$	\dotfill	& $\hatcurLCimp$	\\
~~~$i$ (deg)			\dotfill        & $\hatcurPPi$ \phn 	\\
\sidehead{Spectroscopic parameters}
~~~$K$ (\ms)			\dotfill        & $\hatcurRVK$		\\
~~~$\gamma$ (\kms)		\dotfill 		& $\hatcurRVgamma$	\\
~~~$e$				\dotfill			& $0$ (adopted)		\\
\sidehead{Planetary parameters}
~~~$M_p$ ($\mjup$)		\dotfill		& $\hatcurPPmlong$	\\
~~~$R_p$ ($\rjup$)		\dotfill		& $\hatcurPPrlong$	\\
~~~$C(M_p,R_p)$			\dotfill		& $\hatcurPPmrcorr$	\\
~~~$\rho_p$ (\gcmc)		\dotfill		& $\hatcurPPrho$	\\
~~~$a$ (AU)			\dotfill        	& $\hatcurPParel$	\\
~~~$\log g_p$ (cgs)		\dotfill		& $\hatcurPPlogg$	\\
~~~$T_{\rm eq}$ (K)		\dotfill		& $\hatcurPPteff$	\\
~~~$\Theta$			\dotfill			& $\hatcurPPtheta$
\enddata
\tablenotetext{a}{%
\ensuremath{T_{14}}: total transit duration, time
	between first and last contact; 
\ensuremath{T_{12}=T_{34}}:
	ingress/egress time, time between first and second, or third and fourth
	contact.}
\end{deluxetable}

\begin{figure*}[!ht]
\plottwo{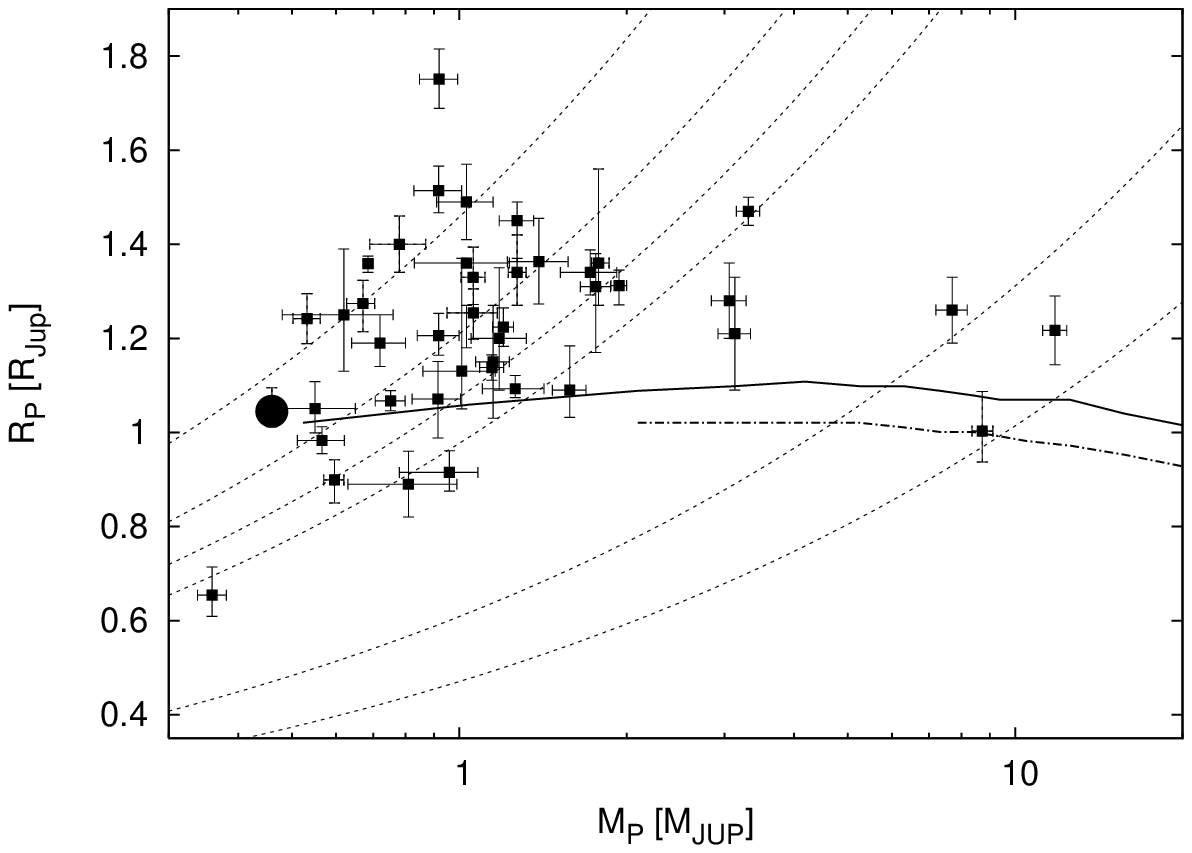}{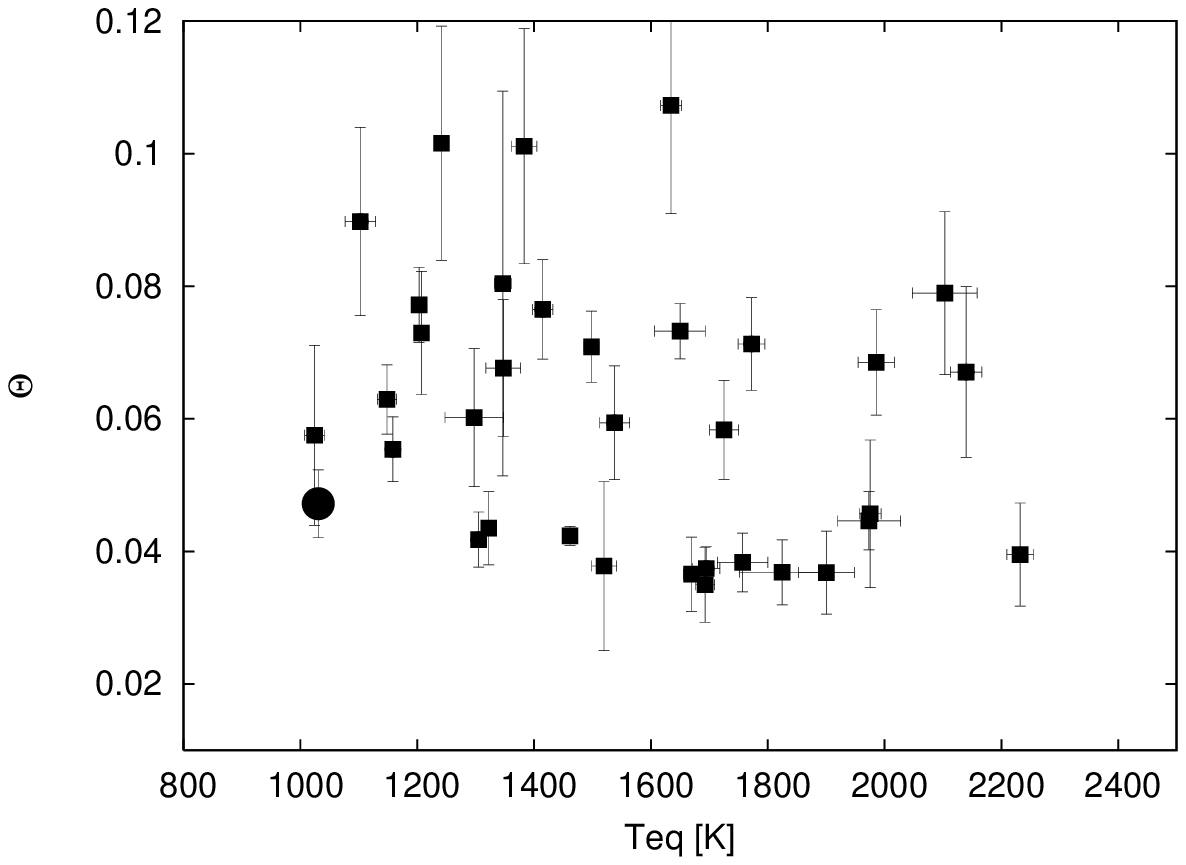}
\caption{
	(Left): Mass--radius diagram of published and uniquely identified
	TEPs. \hatcurb\ is shown as a large filled circle on the left. 
	Overlaid are \citet{baraffe:03} zero insolation planetary
	isochrones for ages of 0.5\,Gyr (upper, solid line) and 5\,Gyr
	(lower dashed-dotted line), respectively, as well as isodensity
	lines for 0.4, 0.7, 1.0, 1.33, 5.5 and 11.9\gcmc (dashed lines). 
	(Right): Equilibrium temperature versus Safronov number
	\citep{hansen:07}.
\label{fig:exomrsaf}}
\end{figure*}

\subsection{Excluding blend scenarios}
\label{sec:blend}

Following \cite{torres:07}, we explored the possibility that the
measured radial velocities are not real, but are instead caused by
distortions in the spectral line profiles due to contamination from a
nearby unresolved eclipsing binary. A bisector analysis based on the
Keck spectra was done as described in earlier HATNet detection papers
(see \S 5 in \cite{bakos:07a}).

The first spectrum in \tabr{rvs} is contaminated, as it was taken at
high airmass, through cloud-cover and in strong moon-light. While the
RV does not seem to be affected, the bisector span is unreliable, thus
we omitted it from the analysis.  We detect no bisector variation in
excess of the measurement uncertainties (see \figr{rvbis} bottom
panel).  We have also tested the significance of the correlation
between the radial velocity and the bisector variations, and these
appear to be negligible.  Therefore, we conclude that the velocity
variations are real, and that the star is orbited by a close-in giant
planet.


\section{Discussion}
\label{sec:discussion}

It is interesting to compare the properties of \hatcurb\ with the other
known TEPs so as to place it in a broader context. This planet falls at
the low-mass end of the current distribution, as shown in
\figr{exomrsaf} (left panel), where we overplot \citet{baraffe:03}
planetary isochrones, which indicate that the radius of \hatcur\ is
broadly consistent with these models. As we noted earlier, \hatcur\ is
formally the smallest mass TEP discovered by transit searches. The even
smaller \hd{149026b} \citep{sato:05} and GJ436b \citep{butler:04} were
discovered by RV searches, and their transits were found later.

We compared our mass and radius to theoretical estimates of
\cite{liu:08} for a 0.5\,\mjup\ body at various orbital distances from
a G2V star. We note that the equivalent semi-major axis (with the same
incident flux) of \hatcurb\ around a solar type star is $a_{rel} =
0.076$. It is also noteworthy that when a detailed comparison is done,
the effects of the environment on the planetary properties are not as
simple as scaling the integrated stellar flux, since the detailed
spectrum of the star (e.g.~UV flux) may also be important.

Based on the models presented by \cite{liu:08}, for
$\dot{E_{h}}/\dot{E}_{ins} = 10^{-6}$ the equilibrium radius of
\hatcurb\ would be $\sim$1.1\,\rjup, where $\dot{E_{h}}$ is the energy
per unit time due to orbital tidal heating or similar internal heating,
and $\dot{E}_{ins}$ is the energy received via insolation. For larger
values of $\dot{E_{h}}/\dot{E}_{ins}$ the expected radius is larger,
and for smaller values it asympotically converges to 1.1\,\rjup.  This
makes us conclude that a small core of approx.~20\mearth\ is required
so that the model values match the observed \hatcurPPrshort\,\rjup\
radius of \hatcurb. This would be also consistent with the
core-mass---stellar metallicity relation proposed by \cite{burrows:07}

When comparing with models of \cite{fortney:08}, we obtain similar
results. The current mass, radius and insolation of \hatcurb\ are
consistent with a 500\,Myr model with a 25\,\mearth\ core mass, or a
4.5\,Gyr coreless pure hydrogen and helium model. It is noted that
low-mass, core-free planets are hard to model, thus our current finding
will hopefully provide a further constraint for theoretical models.

The radiation that \hatcurb\ receives from its host star is $\sim
2.56\cdot10^{8}\ergs$. With the definitions of \cite{fortney:08},
\hatcurb\ belongs to the pL class of planets. There is only one
transiting planet that has a lower mean incident flux: \hd{17156}b
\citep{barbieri:07}, but this planet orbits on a highly eccentric
orbit, with incident flux increasing to over $10^9\ergs$ at periastron.

The other planet with a similarly low incident flux is OGLE-TR-111b
\citep{udalski:02,pont:04}, orbiting an $I=15.55$\,mag star with
$\teffstar = 5040\,K$ (Santos, 2006). \hatcurb\ appears to be a near-by
analog of OGLE-TR-111b in many respects, since their parameters are
very similar (parentheses show those of OGLE-TR-111); the period is
\hatcurLCPshort\,d (4.01\,d), the stellar mass is
\hatcurYYmshort\,\msun\ (0.85\,\msun), the stellar radius is
\hatcurYYrshort (0.83\,\rsun), the luminosity is \hatcurYYlumshort\
(0.4), the metallicity is \hatcurSMEzfeh\ ($0.19\pm0.07$), and the
planetary radius is \hatcurPPrshort\,\mjup\ (1.05). Interestingly, even
the impact parameter of their transits is similar. There is a slight
difference in their masses, with \hatcurb\ being smaller
(\hatcurPPm\,\mjup\ vs.~$0.55\pm0.1\mjup$).
One crucial difference between the two systems is that \hatcur\ is 10
times closer to us, being at \hatcurXdist\,pc vs.~1500\,pc for
OGLE-TR-111, and is more than 4 magnitudes brighter, thus enabling more
detailed follow-up in the near future.

Another interesting observational fact is that the $\Theta =
\hatcurPPtheta$ Safronov number of \hatcurb\ falls fairly close to the
dividing line between the proposed Class I and Class II planets
\citep{hansen:07}. At the low end of the equilibrium temperature range
of the plot (excluding GJ436b), \hatcurb\ seems to be at a point where
the two distributions overlap (see \figr{exomrsaf}, right panel).
Finally, we note that \hatcurb\ strengthens the orbital period
vs.~surface gravity relation \citep{southworth07}, 
falling almost exactly on the linear fit
between these two quantities \citep{torres:08}.


\clearpage
\acknowledgements 

HATNet operations have been funded by NASA grants NNG04GN74G,
NNX08AF23G and SAO IR\&D grants. Work of G.\'A.B.~and J.~Johnson were
supported by the Postdoctoral Fellowship of the NSF Astronomy and
Astrophysics Program (AST-0702843 and AST-0702821, respectively). We
acknowledge partial support also from the Kepler Mission under NASA
Cooperative Agreement NCC2-1390 (D.W.L., PI). G.K.~thanks the Hungarian
Scientific Research Foundation (OTKA) for support through grant
K-60750. This research has made use of Keck telescope time granted
through NOAO (program A285Hr) and NASA (N128Hr).





\end{document}